\documentclass[oneside,a4paper,11pt,shownumbers,manualsort]{article}
\usepackage{latexsym}
\usepackage{euscript}
\usepackage{epsfig,amsmath,amssymb}

\def\be{\begin{equation}}
\def\ee{\end{equation}}
\def\bea{\begin{eqnarray}}
\def\eea{\end{eqnarray}}

\long\def\symbolfootnote[#1]#2{\begingroup%
\def\thefootnote{\fnsymbol{footnote}}\footnote[#1]{#2}\endgroup} 

 \large\normalsize

\begin{document}
%\draft

\begin{center}

{\Large \bf  Cosmic strings in a space-time with positive cosmological constant}

\vspace*{7mm} {Yves Brihaye $^{a),}$ \symbolfootnote[1]{{E-mail:yves.brihaye@umh.ac.be}} and Betti Hartmann $^{b),}$
\symbolfootnote[2]{{E-mail:b.hartmann@jacobs-university.de}}}
%and Eugen Radu $^{b}$}
%\symbolfootnote[2]{E-mail:}
\vspace*{.25cm}

${}^{a)}${\it Facult\'e de Sciences, Universit\'e de Mons-Hainaut, 7000 Mons, Belgium}\\
${}^{b)}${\it School of Engineering and Science, Jacobs University Bremen,
28759 Bremen, Germany}\\

\vspace*{.1cm} 

%\vspace*{.3cm}
\end{center}

\begin{abstract}
We study Abelian strings in a fixed de Sitter background. We find that the
gauge and Higgs fields extend smoothly across the cosmological horizon
and that the string solutions have oscillating scalar fields outside the cosmological horizon
for all currently accepted values of the cosmological constant. If the gauge to Higgs boson mass ratio is small enough, the gauge field
function has a power-like behaviour, while it is oscillating outside the
cosmological horizon if Higgs and gauge boson mass are comparable.
Moreover, we  observe that Abelian strings exist only up to a maximal value of the cosmological
constant and that two branches of solutions exist that meet at this
maximal value. We also construct radially excited solutions that only exist
for non-vanishing values of the cosmological constant and are thus a novel feature
as compared to flat space-time. Considering the effect of the de Sitter
string on the space-time, we observe that the deficit angle increases with increasing
cosmological constant. Lensed objects would thus be separated by a larger angle 
as compared to asymptotically flat space-time.
\end{abstract}

\section{Introduction}
Topological defects are believed to have formed during the phase transitions in the early universe. While magnetic monopoles
and domain walls are catastrophic for the universe since they overclose it, cosmic
strings \cite{strings} were believed to be important for the structure formation for a long time. Recent
Cosmic Microwave background (CMB) measurements however excluded cosmic strings
as seeds for structure formation \cite{cmb}. 

In recent years, cosmic strings have been linked to the fundamental superstrings of string theory and this has boosted renewed interest in these objects.
The low energy limit of certain string theories contain so-called F-
and D-strings, where ``F'' stands for fundamental and ``D'' for Dirichlet.
It was also realized that supersymmetric bound states of F- and D-strings exist, so-called
$(p,q)$-strings \cite{pq}.
The formation of networks of such strings has been discussed in a variety of 
string-inspired, hybrid inflation models \cite{polchinski} and the signatures of 
such networks in the CMB anisotropies have been investigated 
\cite{anisotropies}.

A field theoretical model that contains string-like defects is the Abelian Higgs model
\cite{no}. Abelian Higgs strings are infinitely extended with a core radius inversely proportional to the Higgs boson mass and magnetic flux tubes with radius
inversely proportional to the gauge boson mass.
Recently, two copies of the Abelian Higgs model interacting via a potential term have been discussed
as field-theoretical realizations of $(p,q)$-strings \cite{saffin}.

The gravitational effects of Abelian strings have also been investigated  \cite{string_gravity}.
The main observation is that the space-time far away from the string is conical, i.e. contains
a deficit angle. The consequence of this is that cosmic strings would act as
gravitational lenses, which opens a possibility to detect them indirectly \cite{lenses}.

Since diverse astrophysical observations, e.g. redshift measurements
of type Ia supernovae \cite{sn} lead to the assumption that our universe
is dominated by a form of dark energy, a positive cosmological constant, it is surely of interest
to understand the effects of a positive cosmological constant on cosmic string solutions.
That the effect of a positive cosmological constant is non-trivial has 
been shown e.g. in the study of cosmic string loops, which form unavoidably
in the evolution of cosmic string networks. While cosmic string loops
collapse under their own tension in space-times with Minkowski or Robertson-Walker 
metric, this is not the case for large loops in de Sitter space-time \cite{larsen}.
Thus string loops can survive in space-times with positive cosmological constant.

Moreover, the so-called ``de Sitter/Conformal Field Theory'' (dS/CFT) correspondence \cite{dscft}
suggests a holographic duality between a $d$-dimensional dS space
and a conformal field theory ``living'' on the boundary of dS.

The properties of Abelian strings in de Sitter space-time have been discussed using
analytic tools \cite{dsanalytic} as well as numerical ones \cite{bbh,gm2}.
In \cite{bbh}, a model describing Abelian strings coupled minimally to gravity
including a positive cosmological constant have been studied. While this model describes the
interaction of the matter fields with the gravitational fields properly, the space-time
was assumed to have the same symmetries as the string, namely, it was assumed to be cylindrically symmetric. However, the space-time describing our universe with positive cosmological
constant is genuinely  spherically symmetric for an inertial observer. Thus, it is e.g. not difficult to
study spherically symmetric topological defects such as magnetic monopoles in
a spherically symmetric de Sitter space \cite{bhr}, while it becomes more difficult if
one tries to study objects with symmetry different from spherical symmetry.
It was, however, realized in \cite{gm2,mann} that if one studies strings
in a fixed (Anti-)de Sitter background that the equations describing this situation
become ordinary differential equations if one assumes that the fields of the string
depend only on a specific combination of the spherical coordinates.

Abelian strings have been studied in a background space-time with positive cosmological constant
before \cite{gm2}. 
In this paper, we reinvestigate the solutions given in \cite{gm2}. The authors of
\cite{gm2} have constructed these solutions
(in static coordinates) only inside the cosmological horizon, while we show here that
they extend smoothly across the cosmological horizon. We also
present the asymptotic behaviour of the solutions, which is qualitatively
similar to that of magnetic monopoles in de Sitter space \cite{bhr}.
Our results indicate that for all currently accepted values of the cosmological constant, the
solutions have oscillating scalar fields outside the horizon.
Moreover, we present new, radially excited
solutions that don't exist in the flat space-time limit. 

Our paper is organized as follows: in Section 2, we give the model and equations of motion,
while in Section 3, we present the asymptotic behaviour of the solutions.
In Section 4, we discuss our numerical results and conclude in Section 5.

\section{The model}
We study Abelian strings in a fixed de Sitter background. The metric of the background in static, spherical coordinates (representing the coordinates of
an inertial observer)
can be parametrized as follows:
\begin{equation}
\label{metric}
ds^2=-\left(1-\frac{r^2}{l^2}\right) dt^2 + \left(1-\frac{r^2}{l^2}\right)^{-1} dr^2  +r^2\left(d\theta^2 + \sin^2\theta d\varphi^2\right)
\end{equation}
where $l=\sqrt{3/\Lambda}$ is the de Sitter radius and $\Lambda$ is the (positive) cosmological constant.

The Lagrangian for the Abelian strings reads \cite{no}:
\begin{equation}
{\cal L}=-\frac{1}{4} F_{\mu\nu} F^{\mu\nu}
-(D_{\mu}\phi)^{*} D^{\mu}\phi -\frac{\beta}{4}(\phi^{*}\phi-\eta^2)^2
\end{equation}
with the field strength tensor $F_{\mu\nu}=\partial_{\mu}A_{\nu}-\partial_{\nu}A_{\mu}$ of the $U(1)$ gauge field and the covariant derivative $D_{\mu}\phi=\partial_{\mu}\phi-ieA_{\mu}\phi$ of a with coupling constant $e$ minimally coupled complex scalar field $\phi$. The Lagrangian is invariant under a 
local $U(1)$. When $\phi$ attains a non-vanishing vacuum expectation value $\eta$, the symmetry breaks down from $U(1)$
to $1$. The particle content of the theory is then a massive gauge boson with mass $M_W=e\eta$ and a massive scalar field (the Higgs field) with mass $m_H=\sqrt{2\beta}\eta$.

The Ansatz for the gauge and Higgs fields parametrized in spherical coordinates
$(r,\theta,\varphi)$  reads:
\begin{equation}
A_t=A_r=A_{\theta} =0 \ , \  A_{\varphi}= \frac{1}{e}\left(n -P(r,\theta)\right)\ \ , \ \ 
\phi= \eta f(r,\theta) e^{i n \varphi}
\end{equation}
where $n$ is an integer, which corresponds to the winding number of the string.  

We want to study cylindrical configurations here and thus assume that in the following the matter field functions $P$ and $f$ depend only on the specific combination $r\sin\theta\equiv \rho$.
The partial differential equations then reduce to ordinary differential equations that depend
only on the coordinate $\rho$ and in the limit $l\rightarrow \infty$ correspond to the equations
of the Abelian string \cite{no}.

\subsection{The equations and boundary conditions}
Varying the Lagrangian with respect to the gauge and Higgs fields gives the Euler-Lagrange equations, which here reduce to ordinary differential equations
for the field functions $P$ and $f$. These equations describe a 
cylindrical, string-like configuration in a fixed de Sitter background and read:
\begin{equation}
\label{eqp}
\left(1-\frac{\rho^2}{l^2}\right) P'' =2\eta^2 e^2 P f^2 + \frac{P'}{\rho}\left(1+\frac{2\rho^2}{l^2}\right)
\end{equation}
for the gauge field function and
\begin{equation}
\label{eqf}
\left(1-\frac{\rho^2}{l^2}\right) f''=\frac{\beta}{2}\eta^2 f(f^2-1)-  \frac{f'}{\rho}\left(1-\frac{4\rho^2}{L^2}\right) + \frac{P^2 f}{\rho^2}
\end{equation}
for the Higgs field function, where the prime denotes the derivative with respect to $\rho$.

One can use rescaled coordinates and quantities and define
\begin{equation}
 x=\sqrt{\beta}\eta \rho \ \ , \ \ \alpha=\frac{2M_W^2}{M_H^2}=\frac{e^2}{\beta} \ \ , \ \ L=\sqrt{\beta}\eta l
\end{equation}
The equations then depend only on the parameters $L$ and $\alpha$, where the half of the latter
represents the square of ratio of the gauge boson mass to Higgs boson mass.
Note that with this rescaling, we ``measure'' the cosmological constant in units
of the square of the Higgs boson mass. Furthermore, the case $L\rightarrow \infty$ and
$\alpha=0.5$ corresponds to the self-dual, i.e. BPS limit.

The positive cosmological constant leads to the presence of a cosmological horizon
at $\rho=l$, i.e. $x=L$. Here, we impose boundary conditions at $x=L$ such that
the matter fields are regular at this cosmological horizon. Numerically, we 
first integrate the equations on the interval $x\in [0:L]$ subject to the following
boundary conditions:
\begin{equation}
 P(0)=n  \  , \  f(0)=0 \  , \ \left[2 \alpha x P f^2 + 3 P'\right]_{x=L}=0 \  , \  
 \left[x^2f(f^2-1)+6 x f'+2P^2f\right]_{x=L}=0 \ .
\end{equation}

In a second step, we integrate the equations for $x \in [L,\infty]$
by using as initial conditions the numerical values $P(L),P'(L),f(L),f'(L)$ obtained during the integration for $x\in [0:L]$. We then match the solution for $x\in [0:L]$ and for $x\in [L:\infty]$ at the horizon $x=L$.

The energy density $\epsilon=-T_0^0$ reads:
\begin{equation}
\label{ed}
 \epsilon= \eta^4 \left[  \left(1-\frac{x^2}{L^2}\right) ( f')^2 + \left(1-\frac{x^2}{L^2}\right)\frac{1}{2 \alpha} \frac{(P')^2}{x^2} 
+ \frac{P^2 f^2}{x^2}  + \frac{1}{4} (1-f^2)^2 \right]  \ .
\end{equation}

The inertial mass per unit length inside the cosmological horizon, $M_{in}$, can then be defined by integrating $T_0^0$
over a section of constant $z$, leading to 
\begin{equation}
             M_{in} = 2\pi \eta^2 \int_0^{L} dx \ x \ T_0^0  \ .
\end{equation}

\section{Asymptotic behaviour}
The asymptotic behaviour of the solutions of Eqs.(\ref{eqp}),(\ref{eqf}) plays a major role in the discussion
and depends crucially on the signs of two dimensionless combinations of the mass scales of the theory, namely
on $R_1\equiv 1-8 \alpha L^2$ and $R_2 \equiv 9-4L^2$.
We discuss the different cases separately.
\begin{enumerate}
\item For $R_1 > 0$ and $R_2 >0$, we have
\begin{eqnarray}
\label{Ppower}
        P(x >> 1) &=& P_0 x^c  \ \ , \ \ 
        c = \frac{-1 \pm \sqrt{R_1}}{2}   \ \ , \\ 
\label{fpower}
f(x >> 1) &=& 1 - F_0 x^d  \ \ , \ \ d = \frac{-3 \pm \sqrt{R_2}}{2}
\end{eqnarray}
where $P_0$, $F_0$ are constants to be determined. Note that in contrast to the
case with $\Lambda=0$, the gauge and Higgs field functions decay power-like and
not exponentially.

\item For $R_1 < 0$, $R_2 < 0$, which turns out to be the most relevant case since we expect $L \gg 1$, i.e. the cosmological constant to be much smaller than the square of the Higgs boson
mass from astrophysical
observations, we have instead
\begin{eqnarray}
\label{Poscillating}
        P (x>>1) &=& P_0 x^{-1/2} \sin\left(\sqrt{-R_1/2} \log x + \phi_1\right) \ ,  \\
\label{foscillating}
        f (x >>1) &=& 1 - F_0 x^{-3/2} \sin\left(\sqrt{-R_2/2} \log x + \phi_2\right)
\end{eqnarray}
where $P_0$, $F_0$ and $\phi_1$, $\phi_2$ are constants. We see in particular that both the gauge and Higgs field functions
develop oscillations for $x >>1$.

\item For $R_1 > 0, R_2 <0$ the gauge field behaves like in (\ref{Ppower}) and
the Higgs field like in (\ref{foscillating}).

\item For $R_1 < 0, R_2 > 0$ the gauge field behaves like in (\ref{Poscillating}) and
the Higgs field like in (\ref{fpower}).
\end{enumerate}

%%%%%%%%%%%%%%%%%%%%%%%%%%%%%%%%%%%
\section{Numerical results}
%%%%%%%%%%%%%%%%%%%%%%%%%%%%%%%%%%%%%

Because the equations (\ref{eqp}),({\ref{eqf}}) do not, to our knowledge, admit explicit solutions,   
we have solved them numerically using the ODE solver COLSYS \cite{colsys}. 

Studying the equations numerically, we found that next to the natural deformations
of the standard Abelian Higgs strings (which we call ``fundamental string solutions'' is the following), there exist solutions  for which
the Higgs field function vanishes at some intermediate value of the radial coordinate between the origin and the cosmological horizon. The scalar field function thus develops
nodes. In the following, we will discuss
these two different types of solutions and index them by the number $k$ of nodes. The fundamental
solution thus corresponds to $k=0$.

%%%%%%%%%%%%%%%%%%%%%%%%%%%%%%%%%%%%%%%%%%%%%
\subsection{Fundamental string solutions}
%%%%%%%%%%%%%%%%%%%%%%%%%%%%%%%%%%%%%%%%%%%%%%
First, we have constructed solutions corresponding to $n=1$.

\begin{figure}[!htb]
\centering
\leavevmode\epsfxsize=11.0cm
\epsfbox{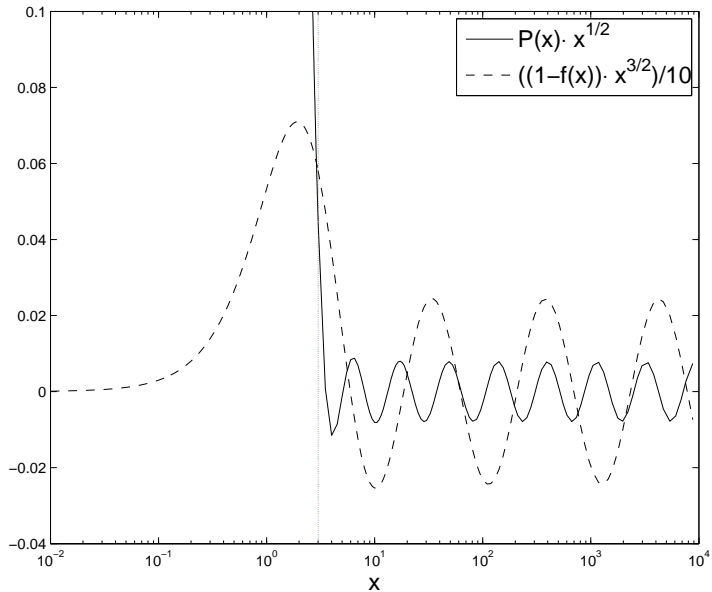}\\
\caption{\label{f1} The  profiles of the 
quantities $P(x)x^{1/2}$ and $((1-f(x))x^{3/2})/10$ are shown for
a de Sitter string with $L=3$ and $\alpha=2$. The localisation of the
horizon at $x=L=3$ is also indicated.  }
\end{figure}

As a first step, we have chosen to find a solution with $R_1<0$ and $R_2<0$ (which
we believe is the physically most relevant case). We have thus chosen
$L=3$, i.e. $R_2=-27$ and $\alpha=2$, i.e. $R_1=-144$. The solution
is shown in Fig.\ref{f1}. In order to see the asymptotic behaviour
predicted in (\ref{Poscillating}) and (\ref{foscillating}), we plot the
quantities $P(x)x^{1/2}$ and $(1-f(x))x^{3/2}$. The oscillations for $x > L$ are then apparent.

\begin{figure}[!htb]
\centering
\leavevmode\epsfxsize=11.0cm
\epsfbox{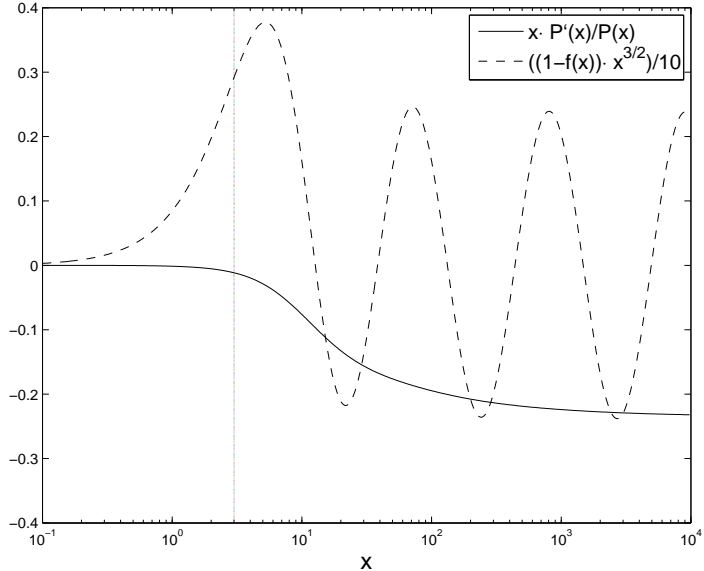}\\
\caption{\label{f2} The profiles of the quantities $xP'(x)/P(x)$ and $(1-f(x))x^{3/2}/10$ are shown for a de Sitter string with $L=3$ and $\alpha=0.01$.
The location of the cosmological horizon at $x=3$ is also indicated.}
\end{figure}

We also present a $n=1$-solution for $R_1 >0$ and $R_2 <0$ in Fig.\ref{f2}. We have chosen 
again $L=3$, but this time $\alpha=0.01$, i.e. $R_1=0.28$.  
The oscillation in the scalar field is apparent when plotting $(1-f(x))x^{3/2}$, while
it is obvious from the plotted quantity $xP'(x)/P(x)$ that the gauge field is behaving power-like as in (\ref{Ppower}). Note that $xP'(x)/P(x)$ tends to a constant
$\sim -0.23$ for $x\rightarrow\infty$. 

We would like to stress that in both cases this correct asymptotic behaviour was NOT imposed as boundary condition, but
was found numerically by imposing the appropriate conditions at the horizon.

\begin{figure}[!htb]
\centering
\leavevmode\epsfxsize=12.0cm
\epsfbox{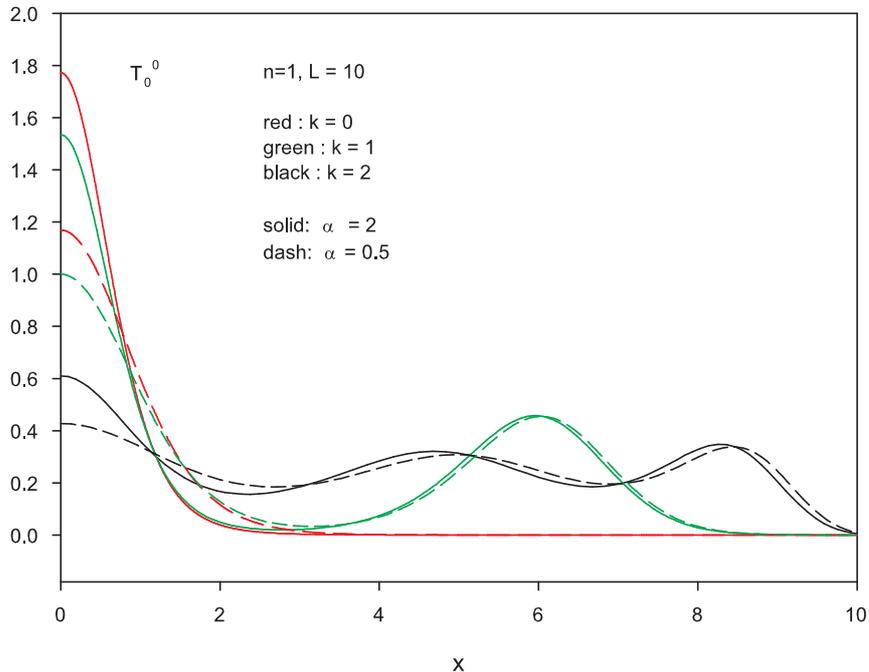}\\
\caption{\label{t00} The energy density $T_0^0$ inside the cosmological horizon
is shown for $k=0$, $k=1$, $k=2$. $k=0$ corresponds
to the fundamental solution, $k=1$ and $k=2$ to the first and second radially excited
solutions, respectively. Here, we have chosen $n=1$, $L=10$, $\alpha=2$ (solid) and $\alpha=0.5$ (dashed), respectively.}
\end{figure}

The energy density $T_0^0$ of a typical de Sitter string solutions with $n=1$, $L=10$, $\alpha=2$ and
$\alpha=0.5$, respectively, is shown in Fig.\ref{t00} (the $k=0$ curves).
Apparently, the energy density is peaked around $x=0$ for the
fundamental string. The corresponding inertial mass per unit length $M_{in}$ for this solution
is $M_{in}/(2\pi\eta^2)\approx 0.75$ for $\alpha=2$ and 
$M_{in}/(2\pi\eta^2)\approx 0.99$ for $\alpha=0.5$, respectively. We observe that when increasing $L$,
the inertial mass increases and reaches the well-known values for $L\rightarrow \infty$, i.e. in the flat space-time limit. Note that $\alpha=0.5$ corresponds to the self-dual limit. We would thus expect
that $M_{in}/(2\pi\eta_2)\rightarrow 1$ for $\alpha=0.5$ and $L\rightarrow \infty$. This is indeed
what we find numerically.

One could then ask whether solutions with a power-like behaviour of the Higgs
field ($R_1 <0$ and $R_2 >0$) or even solutions with a power-like behaviour
of both the gauge and Higgs field ($R_1 >0$ and $R_2 >0$) are possible. 
We will show in the following that solutions of this type do not exist - at least
in our numerical study they don't and we believe that we have constructed
all possible de Sitter string solutions.

Let us explain this in more detail. In order to understand the solutions, we have
studied their domain of existence in the $\alpha$-$L$-plane.
One would expect that some sort of limiting behaviour exists, namely
when $\Lambda\propto 1/L^2$ becomes comparable to the two other mass scales
in the theory, i.e. $M_H^2$ and $M_W^2$.

\begin{figure}[!htb]
\centering
\leavevmode\epsfxsize=11.0cm
\epsfbox{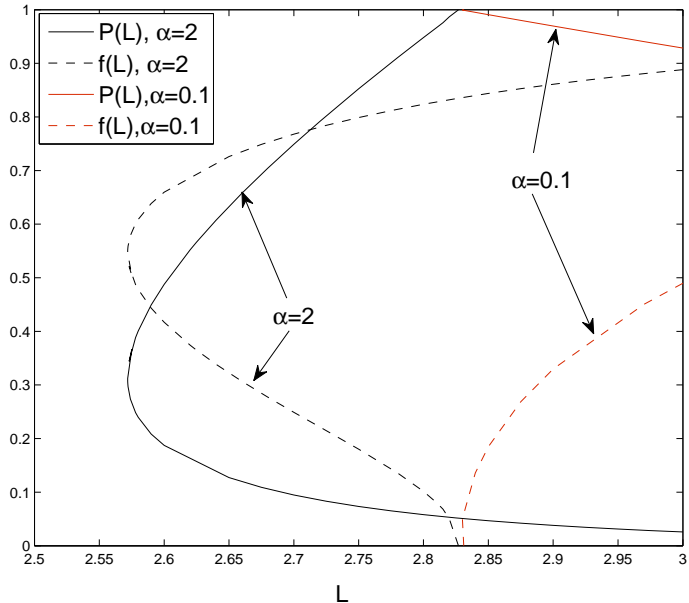}\\
\caption{\label{crit1} The quantities $P(L)$ and $f(L)$ are plotted as functions
of $L$ for $\alpha=2$ (black) and $\alpha=0.1$ (red).}
\end{figure}

\begin{figure}[!htb]
\centering
\leavevmode\epsfxsize=11.0cm
\epsfbox{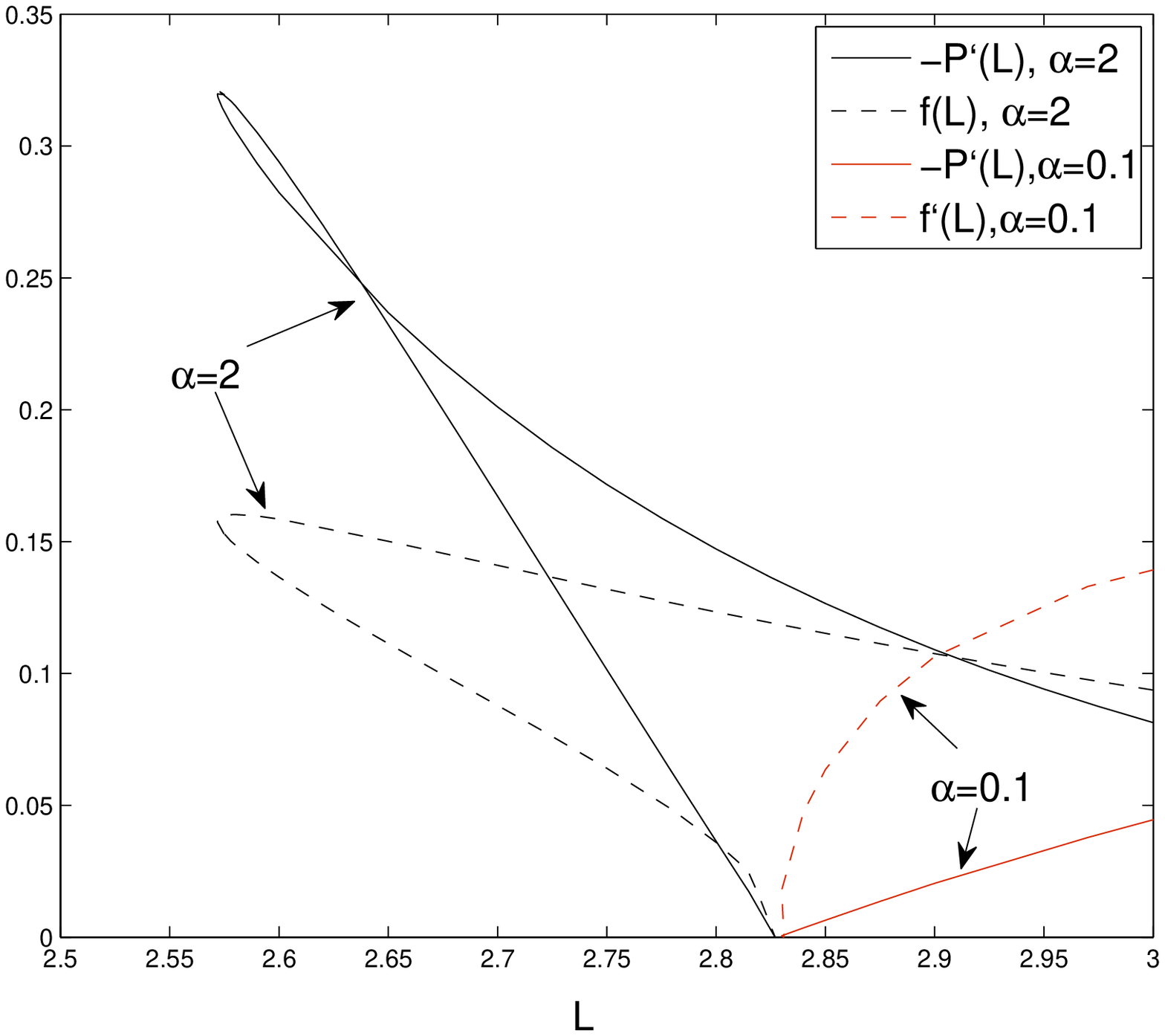}\\
\caption{\label{crit2} The quantities $-P'(L)$ and $f'(L)$ are plotted as functions
of $L$ for $\alpha=2$ (black) and $\alpha=0.1$ (red).}
\end{figure}

Let us first mention that -fixing $\alpha > 0$- we were able to construct solutions which
approach the corresponding well-known string solution in flat space-time for 
$L \to \infty$, i.e. $\Lambda \rightarrow 0$.  Accordingly we find for the values
of the matter field functions and their derivatives at the cosmological horizon $P(L)$, $f(L)$, $P'(L)$, $f'(L)$:
$P(L\rightarrow\infty) \to 0$, $f(L\rightarrow\infty) \to 1$, $P'(L\rightarrow \infty)\to 0$ and $f'(L\rightarrow \infty) \to 0$  - irrespectively of 
the value of $\alpha$. Decreasing the radius $L$, we find a branch of de Sitter strings which extends smoothly for the flat space-time limit with 
$P(L) > 0$, $f(L) < 1$, $P'(L) > 0$ and $f'(L) > 0$. This branch extends all the
way back to a minimal value of the horizon radius at $L=L_{min}(\alpha,n) > 0$. This
is shown in Fig.s \ref{crit1}, \ref{crit2}. E.g. for $n=1$, we find $L_{min} \approx 2.725 $ for $\alpha = 1$ and 
$L_{min} \approx 2.572$ for $\alpha = 2$. $L_{min}$ thus decreases with increasing $\alpha$. The explanation is obvious: when $\alpha$ is increased, the core
radius of the string solution decreases and thus the de Sitter radius can
also be decreased before it becomes comparable to the core radius. 

For large values of $\alpha$, we find that a second branch of de Sitter strings
exists which extends back to a critical value of the horizon
radius $L=L_{cr}$ at which it bifurcates with the trivial solution
$P(x) = 1$, $f(x) =0$. This is shown in Fig.s \ref{crit1},\ref{crit2} for $\alpha=2$.
The existence of the two branches can be explained as follows. Since $L$ is defined
as $L=\sqrt{\beta}\eta l$, the variation of $L$ can either be understood as fixing $l$ and
varying $\sqrt{\beta}\eta$, i.e. the Higgs boson mass, or by fixing $\sqrt{\beta}\eta$ and varying $l$.
The limit $L\rightarrow \infty$ on the first branch corresponds to the flat space-time background with $l\rightarrow \infty$. In flat space-time and $\sqrt{\beta}\eta$ fixed string solutions with
a well defined core radius that is inversely proportional to $\sqrt{\beta}\eta$ exist.
Decreasing $l$, i.e. increasing the cosmological constant, a branch of de Sitter string solutions exists.
These solutions describe strings with a well-defined core radius inside a cosmological horizon.
$l$ can be decreased down to where it becomes comparable to the core radius. This point corresponds to 
the minimal value of $L$, $L_{min}$. From there, a second branch of solutions exists, on which $l$ is kept fixed
while $\sqrt{\beta}\eta$ is varied up to $L_{cr}$.
This works as long as the core radius is larger or comparable to the radius of the
corresponding magnetic flux tube that is given by the inverse of the gauge boson mass.
For smaller values of $\alpha$, i.e. when the radius of the magnetic flux
tube is larger than the core radius, no second branch exists and the branch of
de Sitter solutions bifurcates with the trivial solution at $L=L_{min}=L_{cr}$.
Decreasing $L$ in this case, the cosmological horizon ``sees'' the magnetic flux
tube first, since the core of the string lies within the flux tube. Since the variation
of $L$ can result from the variation of the Higgs boson mass, but not from
the variation of the gauge boson mass at fixed $l$, there is no possibility for a second branch in this case.
Interestingly, our numerical results indicate that $L_{cr}$ depends only slightly on
$\alpha$. We find $L_{cr}\approx 2.83$.

\begin{figure}[!htb]
\centering
\leavevmode\epsfxsize=11.0cm
\epsfbox{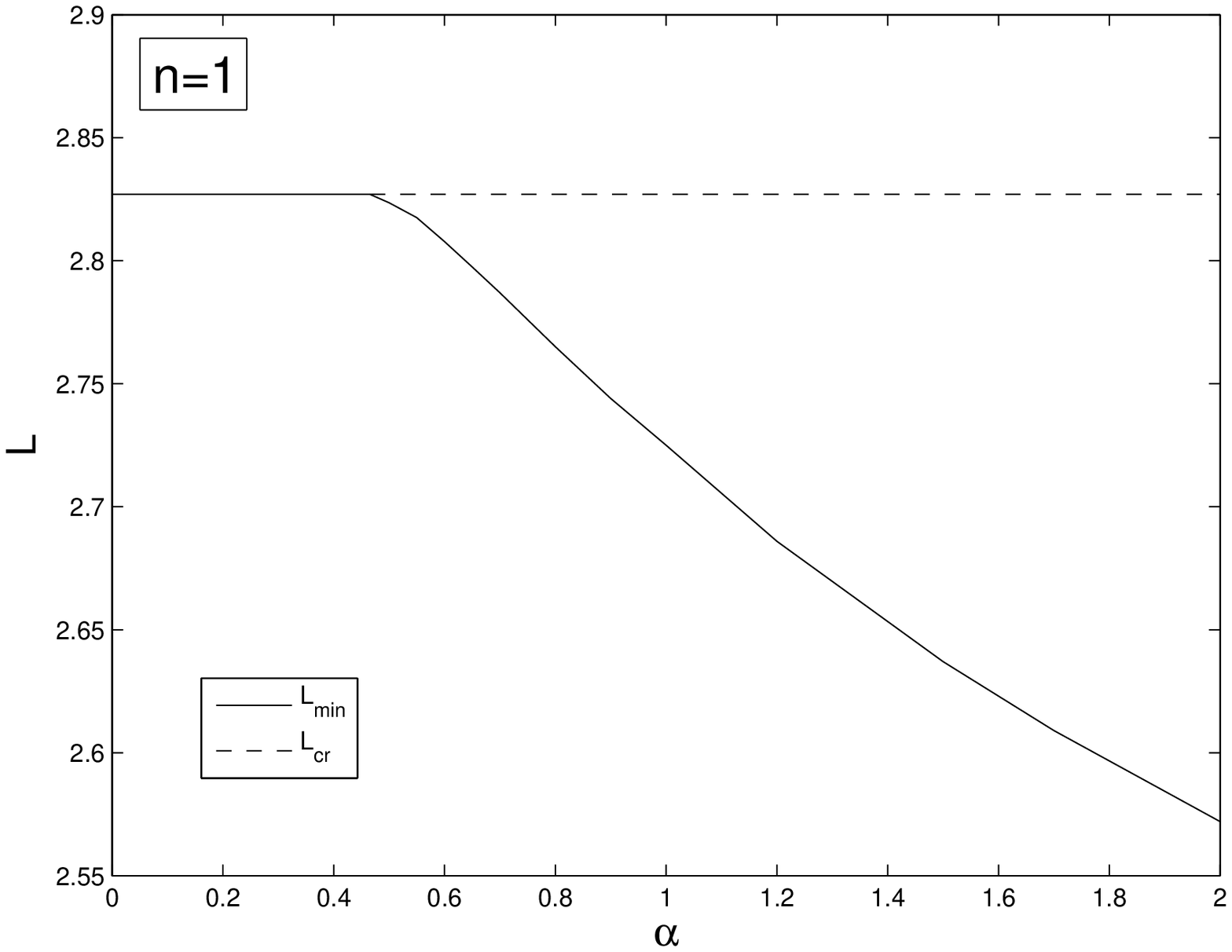}\\
\caption{\label{pattern} The domain of existence of de Sitter strings is
shown in the $\alpha$-$l$-plane. Plotted are the
value of $L_{min}$, the minimal possible value of the de Sitter radius
in dependence on $\alpha$, the square of the ratio of the gauge to Higgs boson mass.
Also shown is the critical value of $l$, $l_{cr}$, where the solutions
become trivial $P(x<\infty)=1$ and $f(x< \infty)=0$. Note that solutions exist only above the solid line.
 }
\end{figure}

In Fig.\ref{pattern}, we show the dependence of $L_{min}$ and $L_{cr}$
on $\alpha$. We find that $L_{min}$ becomes equal to $L_{cr}$ at $\alpha\approx 0.466$.
Note that de Sitter string solutions exist only in the parameter domain above the
solid line. While solutions that fulfill the bound $R_1=1-8\alpha L^2 > 0$, i.e.
have power-like decaying gauge fields are possible, {\bf no} solutions with
$R_2 = 9-4L^2 >0$, i.e. $ L < 3/2$ exist. Thus, all de Sitter strings
have an oscillating Higgs field function outside the cosmological horizon. We believe
that we have constructed all possible de Sitter strings and that no ``isolated'' branches
exist in the $\alpha$-$L$-plane.

\begin{figure}[!htb]
\centering
\leavevmode\epsfxsize=11.0cm
\epsfbox{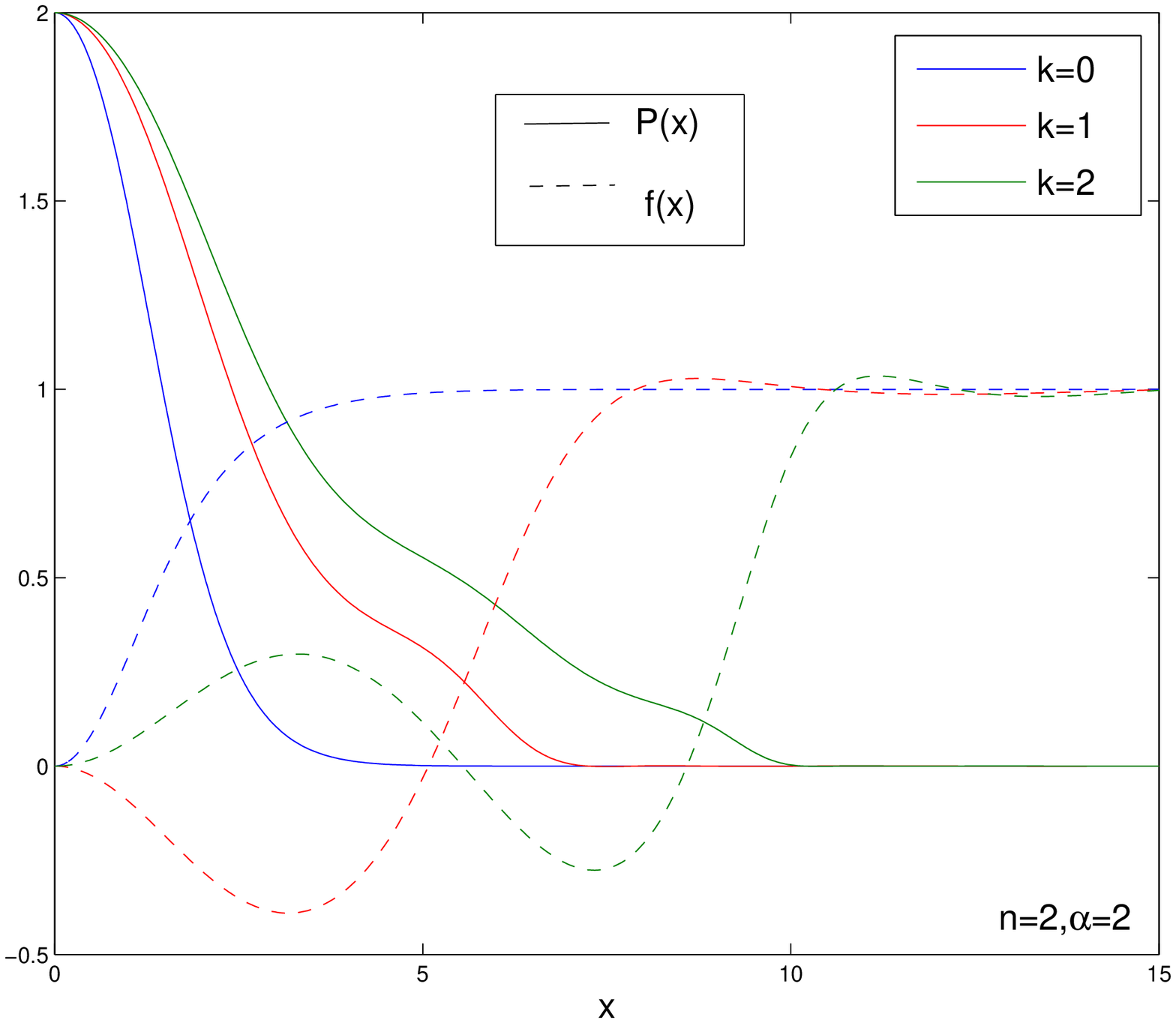}\\
\caption{\label{fig_n2} We show the profiles of the gauge field function $P(x)$ (solid)
and of the Higgs field function $f(x)$ (dashed) of the fundamental ($k=0$, blue), the 1. excited ($k=1$, red) and 2. excited ($k=2$, green)
$n=2$ solution, respectively. Here, $\alpha=2$ and $L=7$ for $k=0$, $k=1$, while $L=9$ for $k=2$.
 }
\end{figure}

We have also studied higher winding solutions. In Fig.\ref{fig_n2}, we present the profiles of a solution for $n=2$ and $\alpha=2$ (blue curves). The study of the dependence of
the solutions on $\alpha$ and $L$ leads to a similar pattern as in the $n=1$ case.
We observe that for fixed $\alpha$ the minimal value of the horizon radius
increases with $n$, e.g. we find $L_{min}(\alpha=2,n=2)\approx 3.460$, while
$L_{min}(\alpha=2,n=1)\approx 2.572$ (see previous discussion). This is related to the fact that the $n=2$ solution has a larger core radius as compared to the $n=1$ solution.

We believe that the qualitative features are similar for $n \ge 3$, this is 
why we don't discuss them here.

%%%%%%%%%%%%%%%%%%%%%%%%%%%%%%%%%%%%%%%%%%
\subsection{Radially excited solutions}
%%%%%%%%%%%%%%%%%%%%%%%%%%%%%%%%%%%%%%%%%%

As mentioned above, our numerical results strongly suggest that 
new types of solutions exist in the presence of a positive cosmological constant.
These solutions are characterized by nodes of the Higgs field function and can be interpreted as radially excited solutions of the fundamental string solutions
discussed in the previous subsection.
Here we present our results for solutions with one and two nodes. We believe that
these are the first members of a tower of solutions labelled by the number of nodes $k \ge 1$ of the function $f(x)$. Note that with this notation the $k=0$ solution
corresponds to the fundamental solution discussed above. 

The comparison of a $k=0$, $k=1$ and $k=2$ solution is given for $n=2$, $\alpha=2$ and $L=7$ for $k=0$, $k=1$, while $L=9$ for $k=2$ in 
Fig.\ref{fig_n2}. We observe that the function $f(x)$ of the $k=1$ solution
first decreases, reaches a minimum and then crosses zero before reaching its asymptotic
value. Note that the value of the radial coordinate $x=x_0$ at which $f(x_0)=0$
($x_0\approx 5.05$ for the solution shown in Fig.\ref{fig_n2}) is smaller than the
corresponding $L$. The gauge field function $P(x)$ remains monotonically decreasing,
but develops a ``shoulder'' in the region of $x_0$. Moreover, the radius of the
core of the excited solution is larger than that of the fundamental solution.
Similarly, the function $f(x)$ of the $k=2$ solution reaches its asymptotic value
after crossing the $x$-axis twice, while the gauge field function $P(x)$ develops
two shoulders at the respective zeros of the function $f(x)$. 
We have not studied the critical behaviour of the $k=2$ solution in detail,
but we believe that it is qualitatively equal to the $k=0$ and $k=1$ cases.

We also show the energy density $T_0^0$ of the $k=1$ and $k=2$ radially excited solutions
in Fig.\ref{t00}. Here $n=1$, $L=10$, $\alpha=2$ and $\alpha=0.5$, respectively.
Clearly, for the excited solutions, $T_0^0$ develops local maxima around the radii coresponding 
to the nodes of the function $f$. 

For $k=1$, we  find $M_{in}/(2\pi\eta^2)\approx 6.91$ for $\alpha=2.0$
and $M_{in}/(2\pi\eta^2)\approx 7.15$ for $\alpha=0.5$, respectively.
As expected, the mass inside the cosmological horizon of the excited
solution is higher than the mass of the fundamental, i.e. $k=0$ solution. 
Equally, one would expect that the mass of the excited solutions with more
than one node of the Higgs field function is even higher.
This is confirmed by our data for $k=2$: we find
$M_{in}/(2\pi\eta^2)\approx 11.21$ for $\alpha=2.0$
and $M_{in}/(2\pi\eta^2)\approx 11.44$ for $\alpha=0.5$, respectively.

\begin{figure}[!htb]
\centering
\leavevmode\epsfxsize=11.0cm
\epsfbox{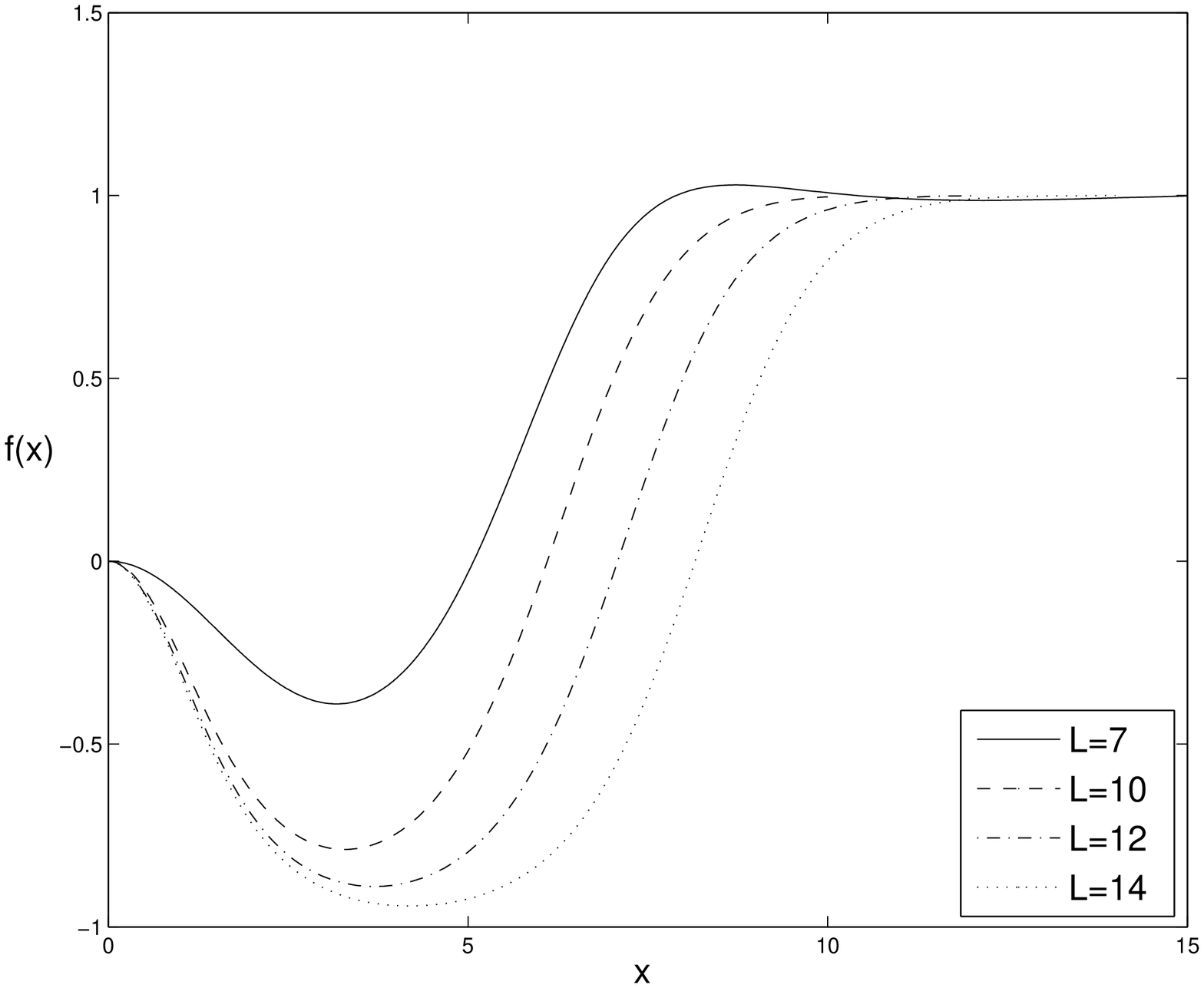}\\
\caption{\label{f_ex} 
 The profiles of the Higgs field function $f(x)$ of the 1.excited solution are shown for $n=2$, $\alpha=2$ and
different values of $L$.}
\end{figure}

Since radially excited solutions don't exist for the flat space-time limit it is natural to study the evolution of the solutions in terms of $L$ and $\alpha$, especially
for $L\rightarrow \infty$. For this, we present the profiles of $f(x)$ for increasing
$L$ in Fig.\ref{f_ex}.

The numerical results given in Fig.\ref{f_ex} suggest that the minimal value of $f(x)$
tends to $-1$ in the limit $L\to\infty$ and that the value of $x_0$ tends to infinity.
Thus for $L\to\infty$, the solution approaches the corresponding Higgs field
function of the Abelian string tending monotonically from $0$ to $-1$. (Note that
normally the Higgs field function of the Abelian string tends from $0$ to $1$,
but that the equations of motion are invariant under $f\to -f$.)

A detailed analysis of the excited solution in the limit $L\to \infty$ is not aimed at in this paper.
Within the accuracy of our numerical results, it seems that, for a sufficiently large $L$,
the function $|f|$ attains a maximum $|f(x_m)|=1$ at
a relatively small value of $x=x_m$ (i.e. with $x_m/L \ll 1$) and that $|f(L)=1|$. In the interval $x \in [x_m,L]$,
we have  $P(x)\sim 0$ while $f(x)$ develops several oscillations. The investigation
of a relation between
the corresponding equation for $f$ and some special function is currently underway.

We have also studied the critical behaviour of the $k=1$ solutions and found a qualitatively similar pattern as for $k=0$ case.

\begin{figure}[!htb]
\centering
\leavevmode\epsfxsize=11.0cm
\epsfbox{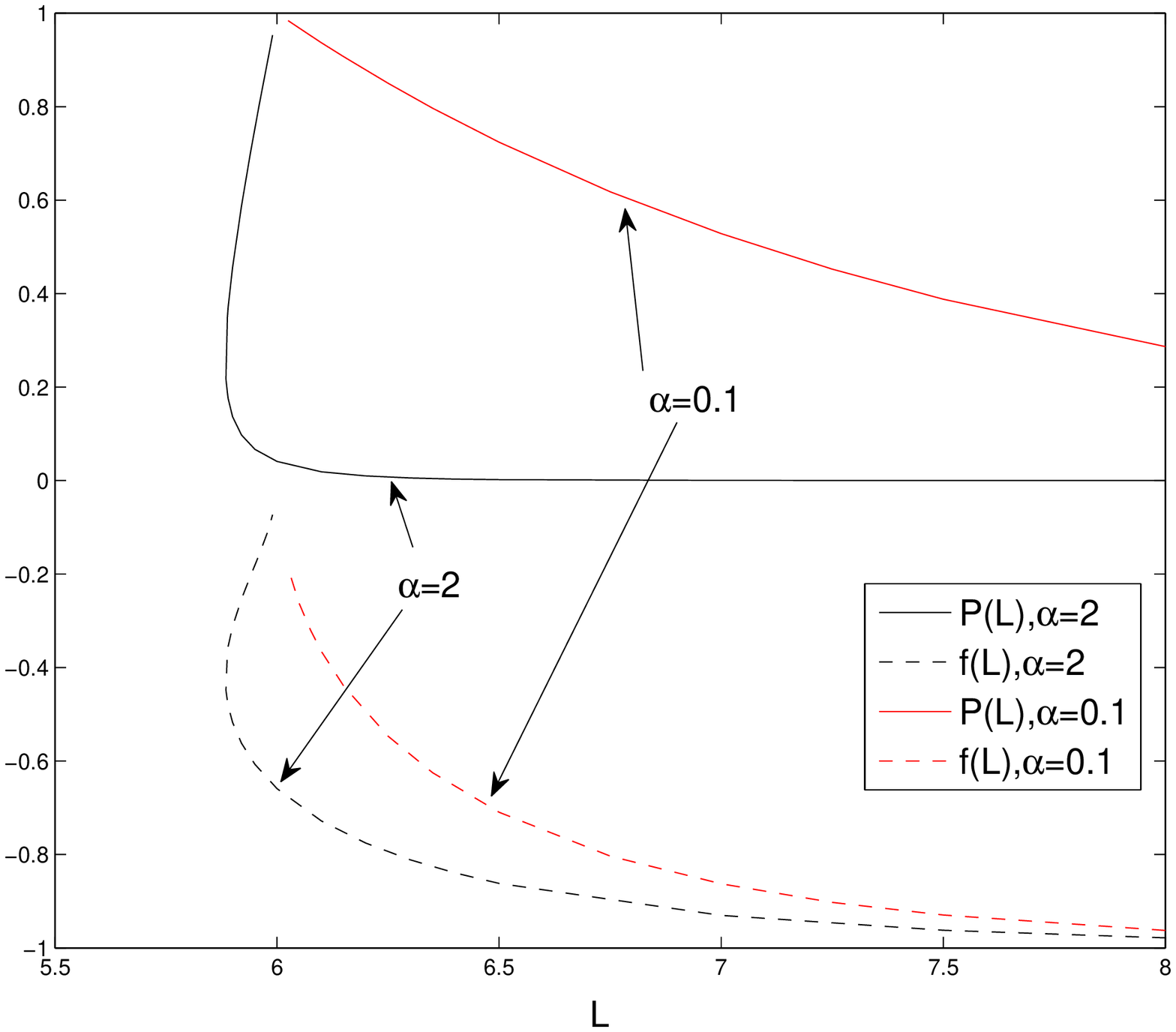}\\
\caption{\label{c_ex} 
 The values of the gauge and Higgs field functions at the horizon, $P(L)$ and $f(L)$, respectively, are shown as functions of $L$ for $\alpha=2$ (black) and $\alpha=0.1$ (red). Here $n=1$ and $k=1$. }
\end{figure}

\begin{figure}[!htb]
\centering
\leavevmode\epsfxsize=11.0cm
\epsfbox{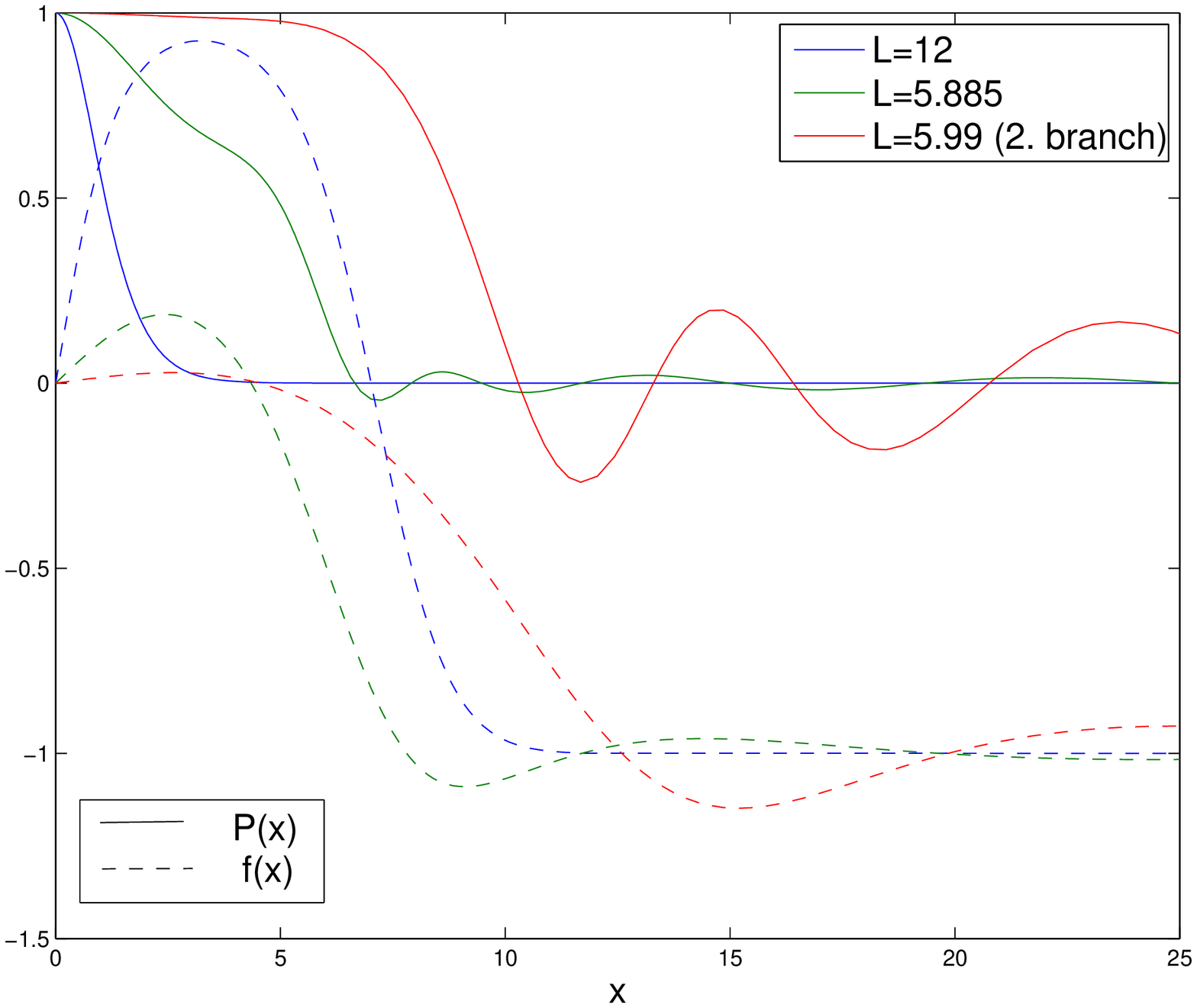}\\
\caption{\label{ex_profiles} 
 The profiles of the gauge and Higgs field functions $P(x)$ and $f(x)$ are shown
for three different values of $L$ and $\alpha=2$, $k=1$, $n=1$.}
\end{figure}

In Fig.\ref{c_ex}, we show the values of the gauge field and Higgs field functions at
the cosmological horizon $P(L)$ and $f(L)$, respectively, as functions of $L$
for $\alpha=2$ and $\alpha=0.1$. Here, we have chosen $n=1$. Again, the radially excited
solutions bifurcate with the trivial solution $P(x)\equiv n$ and $f(x)\equiv 0$ 
at a critical value of $L$: $L_{crit}(\alpha,n)$. We find
$L_{crit}(\alpha,n=1)\approx 6$, where our numerical results
indicate that $L_{crit}$ depends very weakly on $\alpha$.
 
For small values of $\alpha$, the critical value is attained directly by decreasing $L$, while
for larger $\alpha$, a second branch of solutions exists that extends backwards
from a minimal value of the cosmological horizon radius $L_{min}$ with
$L_{min}(\alpha=2,n=1)\approx 5.885$. 

In Fig.\ref{ex_profiles}, we show the profiles of the $k=1$ solution 
for $n=1$, $\alpha=2$ and different values of $L$ along the two branches.
$L=12$ (blue) corresponds to a solution on the first branch and the oscillations
of the functions outside the cosmological horizon are so small that they are not apparent
in the plot. The case $L=5.885$ (green) corresponds to the minimal value of $L$ for this choice
of $n$ and $\alpha$. Here, the amplitude of the oscillations of the fields
 outside the horizon are larger and can be seen in the figure.
Finally, we also present a solution on the second branch, very close to the
critical value of $L$ at $L=5.99$ (red). Here, both $P(x)$ as well as $f(x)$ deviate
only slightly from their values at the origin inside the horizon, while outside the
horizon, they reach their asymptotic values after large amplitude oscillations around these values.
 
\section{Deficit angle}
In all our numerical calculations, we have assumed the de Sitter background to be fixed.
To study the full dynamical space-time is very difficult since the resulting
equations are partial differential equations. In \cite{mann}, an approximation
for weak gravitational fields was used to study the effects of a string
on Anti-de-Sitter space-time. In that case, the Einstein equations can be linearized.
We employ this method here for the de Sitter case. The metric
used reads \cite{mann}
\begin{equation}
 ds^2=\exp(2z/l)\left(-\exp(A)d\hat{t}^2 + d\hat{\rho}^2 + F^2
d\varphi^2\right) + \exp(C)dz^2
\end{equation}
where $A$, $F$ and $C$ are functions of $\hat{\rho}$ and $z$. Introducing
the rescaled coordinate $\hat{x}=\sqrt{\beta}\eta\hat{\rho}$, letting
$z\rightarrow \sqrt{\beta}\eta z$, $t\rightarrow \sqrt{\beta}\eta t$ and 
assuming that the functions
depend only on the combination $x=\hat{x}\exp(z/L)$, the linearized
Einstein equation for $F$ reads
\begin{equation}
\label{manneq}
 \frac{2}{L^2} + \frac{1}{F} \frac{d}{dx}\left((1-\frac{x^2}{L^2})\frac{dF}{dx}\right)=\gamma T^0_0
\end{equation}
where $\gamma=8\pi G$ and $T^0_0$ is the energy-momentum tensor of
the string in the de Sitter background (\ref{ed}).
The deficit angle $\delta$ of the space-time is then given by $\delta=2\pi(1-F'|_{x=x_0})$, where we choose $x_0 < L$.

Outside the core of the string where $T_0^0=0$, we find as solutions to (\ref{manneq}) 
\begin{equation}
    F = c_1 y + c_2 \left(\frac{1}{2} y \log\left(\frac{1+y}{1-y}\right) -1\right) 
\end{equation}
where $y:=x/L$ and $c_1$ and $c_2$ are constants to be fixed by the boundary conditions
of $F$ at the origin. Note that (within the linearized approximation) the function $F(y)$ becomes singular for $y\to 1$, i.e. at the cosmological horizon $x=L$, if $c_2\neq 0$.
Integrating the above equation for our solutions, we find that the function
$F'(y)$ develops a plateau inside the cosmological horizon if $L$ is large enough.
This signals that the space-time inside the cosmological horizon has a
deficit angle. We observe that this deficit angle increase with the decrease of $L$.
Choosing $x_0=L/2$ and integrating (\ref{manneq}) we find the following approximated behaviour of the deficit
angle from our numerics
\begin{equation}
 \delta\approx \gamma M_{in} \left(1 + \frac{8}{3L^2}\right)
\end{equation}

In astrophysical observations it has thus to be taken into account that the presence
of a positive cosmological constant tends to increase the deficit angle as compared
to asymptotically flat-space time. The separation between two lensed objetcs would
thus increase with increasing cosmological constant.

\section{Conclusions}
We have studied Abelian Higgs strings in a fixed de Sitter background. While these
solutions have already been discussed in \cite{gm2}, we find new features
of the solutions here and especially study the behaviour of the matter field
functions outside the cosmological horizon. We find that all possible
de Sitter strings have oscillating Higgs fields outside their horizon.
This observation is important when calculating the mass of these solutions
using the so-called counterterm method \cite{bdm}. For this, we would have to couple
the Abelian Higgs model minimally to gravity. However, in that case, the differential
equations would not reduce to ordinary differential equations (like in our case), but one would
have to solve the ``full'' partial differential equations.
Since it was observed for magnetic monopoles, that the background limit is qualitatively
comparable to the fully coupled case \cite{bhr}, we believe that if we would
couple our model to gravity that the oscillating Higgs fields outside the horizon
would still be a feature of the model. Like in the case of magnetic monopoles
this would then lead to the conclusion that Abelian strings have diverging mass as evaluated at
infinity. The verification of this statement is currently underway and is left for
a future publication.

Moreover, we observe a new feature of the Abelian Higgs model: for non-vanishing cosmological
constant, radially excited solutions exist. Interestingly, the Higgs field function
has nodes in this case. We find that these excited solutions have inertial mass per unit length inside the cosmological
horizon  much larger than the fundamental string solutions. \\
\\
\\
\\

{\bf Acknowlegements} This work has been supported by the ICTS visitors' program of Jacobs University Bremen.
Y. B. thanks the Belgian FNRS for financial support.\\
\\
\\
{\it Note added:} Solutions that are oscillating have also been observed
 in the context of solid state physics \cite{govaerts}.
 Y. B. gratefully acknowledges J. Govaerts for bringing this reference to
 his attention.

\end{document}